\documentclass{modifiedstl}

\usepackage{graphicx}
\usepackage{modifiedams}  
\usepackage{cite}

\begin{document}

\title[Quantum projection filter]{Quantum projection filter for a highly 
nonlinear model in cavity QED}

\author{Ramon van Handel\footnote[3]{E-mail 
address: ramon@its.caltech.edu} and Hideo Mabuchi\footnote[2]{E-mail 
address: hmabuchi@its.caltech.edu}}

\address{Physical Measurement and Control 266-33, California Institute of 
Technology, Pasadena, CA 91125, USA}

\begin{abstract}
Both in classical and quantum stochastic control theory a major role is
played by the filtering equation, which recursively updates the
information state of the system under observation.  Unfortunately, the
theory is plagued by infinite-dimensionality of the information state
which severely limits its practical applicability, except in a few select
cases (e.g.\ the linear Gaussian case.)  One solution proposed in
classical filtering theory is that of the projection filter. In this
scheme, the filter is constrained to evolve in a finite-dimensional family
of densities through orthogonal projection on the tangent space with
respect to the Fisher metric.  Here we apply this approach to the simple
but highly nonlinear quantum model of optical phase bistability of a
stongly coupled two-level atom in an optical cavity.  We observe 
near-optimal performance of the quantum projection filter, demonstrating 
the utility of such an approach.
\end{abstract}



%\maketitle

\section{Introduction}

Over the past decade it has become increasingly clear that feedback
control of quantum systems is essentially a problem of stochastic control
theory with partial observations
\cite{s:belavkin2,s:doherty2,s:VanHandel2005}. In this context, the system
and observations are generally modeled as a pair of It\^o (quantum)
stochastic differential equations.  It is then the goal of the control
engineer to find a feedback control policy, depending on the system state
only through the past history of the observations, that achieves a
particular control objective.

In the case of linear system dynamics and observations and Gaussian
initial conditions, the so-called optimal control problem can be solved
exactly both classically \cite{s:bensoussan} and quantum-mechanically
\cite{s:belavkin2,s:doherty1} provided that a quadratic performance
criterion is chosen.  This means that the control objective is specified
as an optimization problem, where a certain cost function (the performance
criterion) of the system evolution and the control signal is to be
minimized.  The resulting Linear-Quadratic-Gaussian (LQG) control is
widely used in many technological applications.  An important feature of
LQG theory is its {\it separation structure:} the optimal controller
splits up into a part that updates the optimal estimate of the system
state given the observations (the Kalman filter), and an optimal feedback
law which is only a function of the state estimate.

It was originally suggested by Mortensen \cite{s:mortensen} that the
separation structure of LQG control carries over even to the nonlinear
case.  The problem now separates into the nonlinear filtering problem of
finding the optimal estimate of the system statistics given the
observations and the optimal control problem of finding a feedback law,
based only on the filtered estimate, that minimizes some performance
criterion.  The estimate propagated by the filter is often referred to as
the {\it information state} \cite{s:james2} as it contains all information
of the system possessed by the observer.  Unfortunately, nonlinear 
stochastic control is plagued by two rather severe problems.  First, the 
information state is generally infinite-dimensional even for very simple 
nonlinear systems \cite{s:hazewinkelmarcus}. Second, even in the 
finite-dimensional case the nonlinear optimal control problem is generally
intractable.  The latter can sometimes be alleviated by posing a less 
stringent control objective \cite{s:VanHandel2005}. Nonetheless nonlinear 
stochastic control remains an extremely challenging topic, both in the 
classical and quantum mechanical case.

This paper is concerned with the first problem, that of
infinite-dimensionality of the nonlinear information state.  There is no
universal solution to this problem.  The most common (though rather {\it
ad hoc}) approach used by engineers is known as the extended Kalman filter
\cite{s:jacobs}.  In this scheme, the system dynamics is linearized around
the current expected system state, and a Kalman filter based on the linear
approximation is used to propagate the estimate.  However, aside from the 
fact that the method only performs well for nearly linear systems, it is
not clear how it can be applied to quantum models\footnote{
	If the system dynamics can be meaningfully expressed in terms
	of conjugate pairs of observables, one could imagine locally
	linearizing the system Langevin equations to obtain a quantum
	extended Kalman filter.  To our knowledge this has not yet been
	attempted.  However, it is not clear how to do this in e.g.\ 
	atomic systems, where the internal degrees of freedom do not
	obey CCR.
}.

A much more flexible approximation for nonlinear filtering equations was
proposed by Brigo, Hanzon and LeGland \cite{s:brigo1,s:brigo2,s:brigo3},
based on the differential geometric methods of information geometry
\cite{s:amari}.  In this scheme we fix a finite-dimensional family of
densities that are assumed to be good approximations to the information
state.  Using geometric methods the filter is then constrained to evolve
in this family.  The finite-dimensional approximate filter obtained in
this way is known as a projection filter, and often performs extremely 
well when the approximating family is chosen wisely.  Moreover, as this
approximate filter is based on the optimal nonlinear filter, instead of on 
the trajectories of the system state in phase space, it is readily 
extended to the quantum case.  Though by no means a universal solution to 
the filtering problem, we believe that the flexibility and performance of 
this method likely make it widely applicable in the realistic (real-time) 
implementation of quantum filtering theory.

In this paper we apply the projection filtering method to a simple, but
highly nonlinear quantum system: a stongly driven, strongly coupled
two-level atom in a resonant single-mode optical cavity
\cite{s:alsing,s:mabuchi}. The output field of such an experiment exhibits
a randomly switching phase, caused by the atomic spontaneous emission. The
formalism developed by Brigo {\it et al.}\ can be applied directly to this
system if the information state (the conditional density of the atom and
cavity mode) is represented as a $Q$-function \cite{s:mandel}.  
Remarkably, our projection filter shows strong connections to the
classical problem of filtering a random jump process in additive white
noise \cite{s:wonham,s:vellekoop}.

Rather than using a quasiprobability representation, a fully quantum
theory of projection filtering is expressed in terms of finite-dimensional
families of density operators and quantum information geometry
\cite{s:amari}.  We will present the general theory in a future
publication.  Nonetheless there is no theoretical objection to the
approach taken in this paper.  In fact, we observe numerically that the
projection filter for our model has near-optimal performance,
demonstrating the utility of this approach.

This paper is organized as follows.  In section \ref{sec:proj} we
introduce the projection filter and the neccessary elements of information
geometry.  Next, in section \ref{sec:phys}, we introduce the physical
model that we will be using as an example and obtain the associated
filtering equation.  In section \ref{sec:proj2} we obtain the projection
filter for our model.  Finally, in section \ref{sec:num}, we present and
discuss the results of numerical simulations.

\section{Information geometry and the projection filter}
\label{sec:proj}

\subsection{The basic principle of the projection filter}

\begin{figure} 
\includegraphics[width=\textwidth]{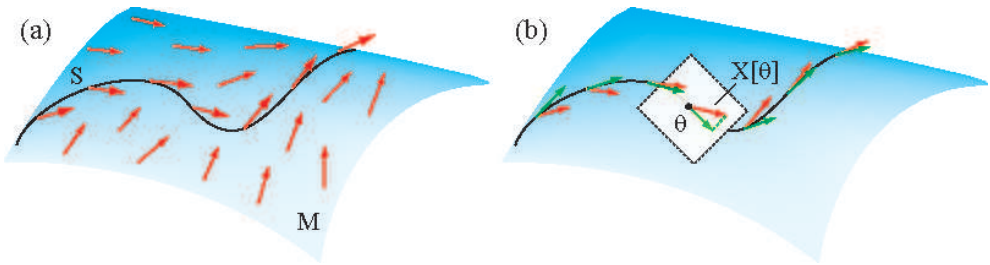} 
\caption{
\label{f:pfilter} Cartoon drawing of the projection filter.  (a) The
infinite-dimensional space of all densities is represented by $M$,
while $S$ is a finite-dimensional submanifold.  The filter, defined
by a (stochastic) differential equation in $M$, flows along a (random) 
vector field on $M$.  The flow is such that the density remains close to 
$S$, but $S$ is not invariant.  (b) To each point $\theta\in S$ the flow 
associates a (random) tangent vector $X[\theta]$ which has components in 
both $T_\theta S$ and its complement.  The projection filter is generated 
by the vector field in $TS$ that, for each $\theta$, is the orthogonal 
projection of $X[\theta]$ onto $T_\theta S$.} 
\end{figure}

The basic idea behind the projection filter is illustrated in Figure
\ref{f:pfilter}.  First, we assume that the information state can be
represented as a probability density, i.e.\ an integrable nonnegative
function on some underlying phase space.  Though this is not always the
case even in classical probability, this is generally a good assumption
for any ``reasonable'' model.  In this paper we will use a well-known
quantum quasiprobability distribution, the $Q$-function \cite{s:mandel},
for this purpose.  The set of all possible densities forms an
infinite-dimensional function space which we will denote by $M$.

We also suppose that the information state is well approximated by
densities in some finite-dimensional subspace $S$ of $M$.  We will assume
that $S$ can be given the structure of a differential manifold, but we do
not require it to be a linear space.  As such we must be careful in what
follows in distinguishing between points in $S$, points in the tangent
bundle $TS$, etc., as in any differential geometric situation.

In general, $S$ will not be an invariant set of the filter; if we start
with a density in $S$, the filter will cause the density to evolve into a
neighborhood of $S$ in $M$.  The idea behind the projection filter is
simply to constrain the optimal filter to remain in $S$.  As $S$ is
finite-dimensional, we can then express the projection filter as a
differential equation in a finite set of local coordinates on $S$.

The optimal filter that propagates the information state of the system is 
given by a stochastic differential equation (SDE) in $M$, and we are 
seeking to express the projection filter as an SDE in $S$.  The precise 
meaning of this statement is a somewhat important point which we will 
return to at the end of this section; for now, we can imagine the filter 
to be an ordinary differential equation that is driven by the 
observations, as follows:
\begin{equation}
\label{eq:vectorf}
	\frac{dp_t}{dt}=X[p_t;Y_t]
\end{equation}
Here $p_t\in M$ is the information state and $Y_t$ is the observation made 
at time $t$.  $X$, then, is an observation-dependent vector field on $M$.

To constrain the filter to evolve in $S$ we must only retain the dynamics
of (\ref{eq:vectorf}) that is parallel to $S$; dynamics perpendicular to
$S$ will move the density into an undesired region.  Mathematically, this
idea is simply implemented if we realize that at each point $\theta\in S$,
$X[\theta;Y]$ will have components both in the tangent space $T_\theta S$
and in its complement $T_\theta S^\perp$.  We can now constrain the vector 
field by orthogonally projecting $X[\theta;Y]$ onto $T_\theta S$ for every 
$\theta\in S$.  The resulting approximate filter, in which only the 
dynamics that leaves $S$ invariant is retained, is the projection filter
\cite{s:brigo1,s:brigo2,s:brigo3}.

Before we can flesh out the details of this scheme we must deal with the 
fact that the filter is not given by a differential equation as in 
(\ref{eq:vectorf}), but by an SDE of the form
\begin{equation}
\label{eq:vectors}
	dp_t=A[p_t]\,dt+B[p_t]\,dY_t
\end{equation}
We would like to think of $A+B\dot Y_t$ as a ``stochastic vector field'' 
so that we can directly apply the scheme discussed above.  The theory of 
stochastic differential equations on manifolds \cite{s:bismut,s:rogersw} 
tells us that we can in fact do this, as long as we interpret 
(\ref{eq:vectors}) as a {\it Stratonovich} SDE
\begin{equation}
\label{eq:stratflt}
	dp_t=A[p_t]\,dt+B[p_t]\circ dY_t
\end{equation}
This is not surprising as, for example, It\^o's rule is incompatible with 
the requirement that the Lie derivative along a vector field is a 
derivation \cite{s:marsden} (in other words, a differential geometric 
transformation rule can only contain first derivatives, and the only 
stochastic integral with this property is the Stratonovich integral.)
Note that usually filtering equations are given in the It\^o form; hence 
we must transform to the Stratonovich form before we can derive the 
projection filter.

\subsection{Information geometry}

In order to perform the key step in the above procedure, the orthogonal
projection, we need an inner product in the tangent space $T_\theta S$.  
A differential manifold is not naturally endowed with an inner product
structure, however, and hence the projection filter is not yet well
defined.  We need to add to the manifold a Riemannian structure
\cite{s:lafontaine}.  In statistics there is a natural way to do this, and
the resulting theory is known as information geometry \cite{s:amari}.

There are different ways of introducing this structure, but perhaps the
easiest treatment is obtained by considering instead of the densities $M$ 
the space of {\it square roots} of densities $M^{1/2}$. The fact that any 
density is integrable guarantees that the square root of any density is 
square integrable; hence $M^{1/2}$ is a subspace of $L^2$, the space of 
square integrable functions, and any vector field on $M^{1/2}$ takes 
values in $L^2$.

Similarly, we consider the manifold $S^{1/2}$, which we will explicitly 
parametrize as
\begin{equation}
   S^{1/2}=\{\sqrt{p(\cdot,\theta)},~~\theta\in\Theta\subset\mathbb{R}^m\}
\end{equation}
That is, $S^{1/2}$ is a finite-dimensional manifold of square roots of 
densities, parametrized by the local coordinates\footnote{
	By writing this, we are assuming that the entire manifold can be
	covered by a single coordinate chart.  Without this assumption
	the description would be more complicated, as then we couldn't
	describe the projection filter using a simple ``extrinsic'' SDE
	in $\mathbb{R}^m$.
	Often we can make our manifold obey this property simply by 
	removing a few points; we will see an example of this later.
} $\theta\in\Theta$.  As $S^{1/2}\subset M^{1/2}$, for any 
$\theta\in\Theta$ the tangent space $T_\theta S^{1/2}$ is the linear 
subspace of $L^2$ given by
\begin{equation}
\label{eq:tthetas}
	T_\theta S^{1/2}=\mbox{Span}\left[
		\frac{\partial\sqrt{p(\cdot,\theta)}}{\partial\theta^1},
		\cdots,
		\frac{\partial\sqrt{p(\cdot,\theta)}}{\partial\theta^m}
	\right]\subset L^2
\end{equation}
The reason for working with square roots of densities is that this gives 
a natural inner product in the tangent space, which is simply the standard 
$L^2$-inner product.  In particular, we can calculate the associated 
metric tensor in the basis of (\ref{eq:tthetas}):
\begin{equation}
\label{eq:fisher}
	\left\langle
	\frac{\partial\sqrt{p(\cdot,\theta)}}{\partial\theta^i},
	\frac{\partial\sqrt{p(\cdot,\theta)}}{\partial\theta^j}
	\right\rangle=
	\int 
	\frac{\partial\sqrt{p(x,\theta)}}{\partial\theta^i}
	\frac{\partial\sqrt{p(x,\theta)}}{\partial\theta^j}dx=
	\frac{1}{4}g_{ij}(\theta)
\end{equation}
Up to a factor of $1/4$, this is the well-known Fisher information matrix
$g_{ij}(\theta)$.

We are now in the position to define what we mean by orthogonal projection 
of a vector field on $M$ onto $TS$.  At each $\theta$, the orthogonal 
projection is
\begin{equation}
\label{eq:orthog}
	\Pi_\theta X[\theta]=\sum_{i=1}^m\sum_{j=1}^m
		4g^{ij}(\theta)
	\left\langle X[\theta],
	\frac{\partial\sqrt{p(\cdot,\theta)}}{\partial\theta^j}
	\right\rangle
	\frac{\partial\sqrt{p(\cdot,\theta)}}{\partial\theta^i},
\end{equation}
where we have used the inverse Fisher information matrix $g^{ij}(\theta)$ 
to account for the fact that the basis of (\ref{eq:tthetas}) is not 
orthogonal.  This is the main result that is needed to obtain projection 
filters.

\subsection{Orthogonal projection of a Stratonovich filter}

Let us now discuss how to perform orthogonal projection onto a 
finite-dimensional manifold $S$ for the very general form 
(\ref{eq:stratflt}) of a filtering equation.  We begin by converting the 
equation to the square root form; this gives
\begin{equation}
	d\sqrt{p_t}=\frac{1}{2\sqrt{p_t}}A[p_t]\,dt
			+\frac{1}{2\sqrt{p_t}}B[p_t]\circ dY_t
\end{equation}
We now constrain the filter to evolve on $S^{1/2}$ through orthogonal 
projection:
\begin{equation}
\label{eq:pfiltfll}
	d\sqrt{p(\cdot,\theta_t)}=
		\Pi_{\theta_t}
		\frac{
			A[p(\cdot,\theta_t)]
		}{2\sqrt{p(\cdot,\theta_t)}}\,dt
		+\Pi_{\theta_t}
		\frac{
			B[p(\cdot,\theta_t)]
		}{2\sqrt{p(\cdot,\theta_t)}}\circ dY_t
\end{equation}
This is just a finite-dimensional SDE for the parameters $\theta_t$.  To 
convert the expression explicitly into this form, note that by the 
Stratonovich transformation rule
\begin{equation}
	d\sqrt{p(\cdot,\theta_t)}=
	\sum_i\frac{\partial\sqrt{p(\cdot,\theta_t)}}{\partial\theta_t^i}
	\circ d\theta_t^i
\end{equation}
Comparing with (\ref{eq:orthog}) and (\ref{eq:pfiltfll}), we find that
\begin{equation}
\label{eq:projf}
	d\theta_t^i=
		\left\langle 
		\frac{
			A[p(\cdot,\theta_t)]
		}{p(\cdot,\theta_t)},\Lambda_t^i(\cdot,\theta_t)
		\right\rangle dt + \left\langle 
		\frac{
			B[p(\cdot,\theta_t)]
		}{p(\cdot,\theta_t)},\Lambda_t^i(\cdot,\theta_t)
		\right\rangle
		\circ dY_t
\end{equation}
where
\begin{equation}
\label{eq:projf2}
	\Lambda_t^i(\cdot,\theta_t)=
	\sum_{j=1}^m
	g^{ij}(\theta_t)\frac{\partial p(\cdot,\theta_t)}{\partial\theta^j_t}
\end{equation}
Equations (\ref{eq:projf}), (\ref{eq:projf2}) and (\ref{eq:fisher}) can be 
used to directly calculate the projection filter for a wide range of 
models.

\section{The physical model and the quantum filter}
\label{sec:phys}

\subsection{The Jaynes-Cummings model in the strong driving limit}
\label{sec:jc1}

\begin{figure} 
\includegraphics[width=\textwidth]{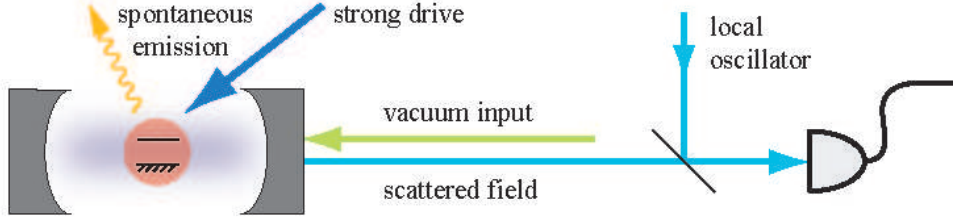} 
\caption{
\label{f:jaynes} Schematic of the experimental setup that corresponds to 
our model.  A strongly coupled two-level atom in a resonant, single mode 
cavity is strongly driven by a resonant driving laser, and spontaneously 
emits in all directions.  One of the cavity mirrors is leaky, and the 
atomic dynamics is observed through homodyne detection of the 
electromagnetic field in this forward mode.}
\end{figure}

We consider the following physical system, shown in Figure \ref{f:jaynes}.
A two-level atom is strongly coupled to the mode of a single-mode cavity.  
The cavity mode and the atomic frequency are resonant.  The atom is
strongly driven on resonance by a laser, and spontaneously emits in all
directions.  A forward mode of the electromagnetic field outside the
cavity, initially in the vacuum state, scatters off one of the cavity
mirrors.  By making this mirror slightly leaky, we extract information
from the system into the external field.  Homodyne detection of the
forward mode then yields information about the atom and cavity.

The goal of this section is to model this physical system as a pair of 
It\^o quantum stochastic differential equations, one of which describes 
the atom-cavity evolution and one describing the homodyne observations.  
To this end, we begin by writing down the full Hamiltonian for the system:
\begin{equation}
	H=H_0+H_d+H_{\rm JC}+H_f+H_s
\end{equation}
Here $H_0$ is the free Hamiltonian
\begin{equation}
	H_0=
	\hbar\omega_0a^\dag a+
	\frac{\hbar\omega_0}{2}\sigma_z
	+\int_0^\infty d\omega\,\hbar\omega (b_f^\dag(\omega)b_f(\omega)
		+b_s^\dag(\omega)b_s(\omega)),
\end{equation}
$H_d$ is the drive Hamiltonian
\begin{equation}
	H_d=i\hbar(\mathcal{E}/2)
		(e^{i\omega_0t}\sigma-e^{-i\omega_0t}\sigma^\dag),
\end{equation}
$H_{\rm JC}$ is the well-known Jaynes-Cummings Hamiltonian
\begin{equation}
	H_{\rm JC}=i\hbar g(a^\dag\sigma-a\sigma^\dag),
\end{equation}
and $H_f$ and $H_s$ are the dipole couplings to the forward and 
spontaneous emission field modes outside the cavity (see e.g.\ 
\cite{s:gardinercollett})
\begin{eqnarray}
	H_f=\hbar\int_0^\infty d\omega\,
	\kappa_f(\omega)[ab_f^\dag(\omega)+a^\dag b_f(\omega)] \\
	H_s=\hbar\int_0^\infty d\omega\,
	\kappa_s(\omega)[\sigma b_s^\dag(\omega)+\sigma^\dag b_s(\omega)]
\end{eqnarray}
Here $\sigma=|g\rangle\langle e|$ is the atomic lowering 
operator, $\sigma_z=[\sigma^\dag,\sigma]$, $a$ is the cavity mode lowering 
operator (we will also use $x=a^\dag+a$ and $y=i(a^\dag-a)$), and 
$b_f(\omega)$ and $b_s(\omega)$ are the annihilators of the forward and 
spontaneous emission modes, respectively.  The resonant frequency of the 
atom, drive and cavity mode is denoted by $\omega_0$, $\mathcal{E}$ is the 
drive strength, $g$ is the atom-cavity coupling strength, and 
$\kappa_f(\omega)$ and $\kappa_s(\omega)$ determine the 
frequency-dependent coupling to the external field modes.

We will assume that $\omega_0\gg\mathcal{E}\gg g>\kappa_f,\kappa_s$.
Following \cite{s:mabuchi}, let us switch to the interaction picture with 
respect to $H_0+H_d$.  We obtain the interaction Hamiltonian
\begin{eqnarray}
	\frac{H_{\rm I}}{\hbar}=
	ig[a^\dag\sigma(t)-a\sigma^\dag(t)]
	+\int_0^\infty d\omega\,
		\kappa_s(\omega)[\sigma(t)b_s^\dag(\omega)
			e^{i(\omega-\omega_0)t}+\mbox{h.c.}]
	\nonumber\\
\label{eq:interaction}
	\phantom{\frac{H_{\rm I}}{\hbar}=ig[a^\dag\sigma(t)-a\sigma^\dag(t)]}
	+\int_0^\infty d\omega\,
		\kappa_f(\omega)[ab_f^\dag(\omega)e^{i(\omega-\omega_0)t}
			+\mbox{h.c.}] 
\end{eqnarray}
where we have defined
\begin{eqnarray}
	|\pm\rangle = 2^{-1/2}(|g\rangle\mp i|e\rangle), ~~~~
	\mu=|-\rangle\langle+|, ~~~~ \mu_z=[\mu^\dag,\mu], \\
	\sigma(t)=(-i/2)(\mu e^{-i\mathcal{E}t}+\mu_z-\mu^\dag e^{i\mathcal{E}t})
\end{eqnarray}
There are two time scales in the Hamiltonian (\ref{eq:interaction}), which 
we will consider separately.  The first term evolves on the slow time 
scale of the atomic evolution.  As $\mathcal{E}$ is very large compared to 
the atomic time scale, we make the rotating wave approximation by dropping 
the rapidly oscillating terms.

The remaining terms in (\ref{eq:interaction}) correspond to the fast time 
scale of interaction with the external electromagnetic field.  We cannot 
use the rotating wave approximation for these terms, as the external 
fields are broadband and thus have modes that respond on the fast time 
scale.  Instead, we make the weak coupling (Markov) approximation for 
these terms; this results in the following white noise Hamiltonian:
\begin{eqnarray}
\label{eq:whiten}
	\tilde H_{\rm I}=
	\hbar(g/2)\mu_zx
	+\hbar\sqrt{2\kappa}\,[a\dot B_f^\dag(t)+\mbox{h.c.}] 
	+i\hbar[
		\sqrt{\gamma_{+}/2}\,\mu\dot B_{s,+}^\dag(t)
	\nonumber\\
	\phantom{\tilde H_{\rm I}=\hbar(g/2)\mu_zx+}
		+\sqrt{\gamma_{z}/2}\,\mu_z\dot B_{s,z}^\dag(t)+
		\sqrt{\gamma_{-}/2}\,\mu^\dag\dot B_{s,-}^\dag(t)-\mbox{h.c.}]
\end{eqnarray}
Here $B_f^\dag$, $B_{s,+}^\dag$, $\dot B_{s,z}^\dag$ and 
$\dot B_{s,-}^\dag$ are independent quantum white noises corresponding, 
respectively, to the forward channel and the three spontaneous emission 
channels at $\omega=\omega_0+\mathcal{E}$, $\omega_0$, and 
$\omega_0-\mathcal{E}$ (the upper, middle and lower peaks of the Mollow 
triplet.)  

We refer to \cite{s:gardinercollett} for a discussion of the white noise
approximation.  Care must be taken to assign an independent white noise to
each frequency component of $\sigma(t)$ (e.g.\ section III.E of
\cite{s:gardinercollett}); each frequency probes a different subset of the
modes $b(\omega)$ and hence ``sees'' a different noise.  In the weak
coupling limit these noises are in fact white and independent.  For a more
rigorous approach to the white noise limit see \cite{s:accardi,s:gough}.  
Using the latter approach we can explicitly calculate
$\kappa=\pi\kappa_f(\omega_0)^2$, $\gamma_z=\pi\kappa_s(\omega_0)^2$, and
$\gamma_\pm=\pi\kappa_s(\omega_0\pm\mathcal{E})^2$.  For simplicity, we
will assume that approximately $\gamma_{z,+,-}=\gamma$.

The white noise Hamiltonian (\ref{eq:whiten}) by itself is not well 
defined.  However, we can give rigorous meaning to the equation
\begin{equation}
	\frac{dU_t}{dt}=-\frac{i}{\hbar}\tilde H_{\rm I}U_t
\end{equation}
if we interpret it as a {\it Stratonovich} quantum stochastic differential 
equation \cite{s:gardinercollett,s:gough}.  After conversion to the It\^o 
form, this equation reads
\begin{eqnarray}
	dU_t=[
		\sqrt{\gamma/2}\,(\mu\,dB_{s,+}^\dag(t)
			+\mu_z\,dB_{s,z}^\dag(t)
			+\mu^\dag\,dB_{s,-}^\dag(t)-\mbox{h.c.})
	\nonumber\\
	\phantom{dU_t=[\sqrt{\gamma/2}}
		-i\sqrt{2\kappa}\,(a\,dB_f^\dag(t)+\mbox{h.c.})
	\nonumber\\
\label{eq:sys}
	\phantom{dU_t=[\sqrt{\gamma/2}}
		-\kappa a^\dag a\,dt
		-(\gamma/2)\,dt
		-i(g/2)\mu_z x\,dt
	]\,U_t
\end{eqnarray}

Let us now turn to the homodyne observation of the field.  The homodyne
detector measures a quadrature of the forward channel after it has
scattered off the cavity.  We will choose the quadrature
$B_f(t)+{B_f}^\dag(t)$; the observation process is then
$Y(t)=U_t^\dag(B_f(t)+{B_f}^\dag(t))U_t$ (i.e., the photocurrent is
$I(t)=dY(t)/dt$.)  Using the quantum It\^o rules 
\cite{s:gardinercollett,s:hudpar}, we easily find the differential form of 
this expression:
\begin{equation}
\label{eq:rawobs}
	dY(t) = \sqrt{2\kappa}\,U_t^\dag yU_t\,dt+
			dB_f(t)+dB_f^\dag(t)
\end{equation}
We can slightly extend our observation model to account for technical 
noise, detector inefficiency, etc.  To model such effects, we add to 
(\ref{eq:rawobs}) an independent corrupting noise $\propto dC(t)+dC^\dag(t)=
dV(t)$.  It is customary in the quantum optics literature to normalize 
$Y(t)$ so that $dY(t)^2=dt$.  In terms of the detection efficiency 
$\eta\in (0,1]$
\begin{equation}
\label{eq:obs}
	dY(t) = \sqrt{2\kappa\eta}\,U_t^\dag yU_t\,dt
		+\sqrt{\eta}\,[dB_f(t)+dB_f^\dag(t)]
		+\sqrt{1-\eta}\,dV(t)
\end{equation}
We will take the It\^o equations (\ref{eq:sys}), (\ref{eq:obs}) as 
our model for the system-observation pair.

\subsection{The quantum filter}
\label{sec:jc2}

Now that we have a model for the system and the observation process, we 
can calculate the optimal filter.  The derivation of the filtering 
equation is beyond the scope of this paper; for various approaches, see 
\cite{s:VanHandel2005,s:belavkin,s:belavkz1,s:belavkz2,s:bouten}.  We will 
attempt, however, through a simple finite-dimensional analogy, to explain 
our interpretation of the filtering equation, as it is not entirely the 
same as the interpretation that is often found in the physics literature 
(e.g.\ \cite{s:wisemanf}).

The optimal filter propagates the information state, which determines our 
best estimate of every system observable given the observations we have 
made.  In our model, every system observable can be represented as a 
self-adjoint operator $X$ that lives on the atom-cavity Hilbert space; as 
we are working in the Heisenberg picture, this observable at time $t$ is 
given by $j_t(X)=U_t^\dag XU_t$.  We must now define what we mean by an 
estimate of an observable.

The idea behind the concept of estimation is that we have made some
observation, and given the outcome of this observation we wish to make a
guess as to the outcome of a different observable that we haven't
measured. That is, the estimate of an observable $X$ given an observation
of $Y$ is some function $f(Y)$ whose outcome represents our best guess of 
$X$.  To find the {\it best} estimate we must specify some cost function
$\mathcal{C}[f]$ to optimize; the function that minimizes $\mathcal{C}$ is 
then by definition the optimal estimate.

The most commonly used estimator is one that minimizes the mean-square 
error
\begin{equation}
\label{eq:meansq}
	\mathcal{C}[f]=\langle(X-f(Y))^2\rangle
\end{equation}
The observable $f(Y)$ that minimizes this cost is called the {\it 
conditional expectation} $\mathcal{E}(X|Y)$ of $X$ given $Y$.
We will use the conditional expectation as our information state 
throughout this paper.  However, note that if we had chosen a different 
cost we would obtain a different information state and filter.  There is 
nothing inherently superior about the choice (\ref{eq:meansq}); in 
fact, it is sometimes advantageous to choose a different estimator with 
e.g.\ improved robustness properties \cite{s:james2,s:james3}.

To understand how the conditional expectation relates to familiar notions 
from quantum theory, we will demonstrate the procedure using a pair of 
finite-dimensional observables \cite{s:maassen}.  Let $X$ and $Y$ be two 
$n$-dimensional observables, $n<\infty$, and let $Y$ have $m$ distinct 
eigenvalues $y_i$.  Then $Y$ can be decomposed as
\begin{equation}
	Y=\sum_{i=1}^m y_iP_i
\end{equation}
where $P_i$ is the projection operator onto the eigenspace corresponding 
to $y_i$.  Clearly any function of $Y$ is a linear combination of $P_i$, 
and vice versa.  Hence we identify the set of all observables that are 
functions of $Y$ with the span of $\{P_i\}$.  The conditional expectation
$\mathcal{E}(X|Y)$ is then the element of this set that minimizes the cost 
(\ref{eq:meansq}).

To find this element, we use the following trick.  The expression $\langle 
X^\dag Y\rangle$ defines an inner product on the set of $n\times n$ 
complex matrices\footnote{
	We assume for simplicity that the expectation map 
	$\langle\cdot\rangle$ is faithful \cite{s:maassen}.  If this is 
	not the case, then the conditional expectation is not unique.
	However, all versions of $\mathcal{E}(\cdot|\cdot)$ are
	equivalent in the sense that the difference between two versions
	takes nonzero values with zero probability.
}.  Using this inner product, we orthogonally project $X$ onto the linear 
space spanned by $\{P_i\}$.  This gives
\begin{equation}
\label{eq:qcondex}
	P_YX=\sum_{i=1}^m\frac{\langle P_iX\rangle}{\langle P_i\rangle}P_i
\end{equation}
It is a well known fact that the orthogonal projection of some vector 
$v$ onto a linear subspace $W$ with respect to any inner product $(a,b)$ 
gives the element $w\in W$ that minimizes the quantity $((v-w),(v-w))$ 
\cite{s:naylor}.  In our case, this means that $P_YX$ minimizes $\langle
(X-f^*(Y))(X-f(Y))\rangle$.  $f(Y)$ is only an observable, however, if $f$ 
is real, in which case we see that $P_YX$ is precisely the conditional 
expectation $\mathcal{E}(X|Y)$.  Note that the orthogonal projection 
$P_YX$ will always be self-adjoint if $X$ and $Y$ commute.

Remarkably, when $X$ and $Y$ commute, the expression (\ref{eq:qcondex}) is 
equivalent to the traditional projection postulate.  To see this, note 
that if we observe $Y=y_i$ then $P_YX$ takes the value 
$\langle P_iX\rangle/\langle P_i\rangle=
\langle P_iXP_i\rangle/\langle P_i\rangle$, which is exactly the 
expectation of $X$ with respect to the initial state projected onto the 
eigenspace of $y_i$.  The situation is somewhat ambiguous for noncommuting 
$X$ and $Y$, and we will simply refrain from defining the conditional 
expectation $\mathcal{E}(X|Y)$ when $[X,Y]\ne 0$.

The quantum filter determines the best estimate of every system observable
given the observations; i.e., it propagates
$\pi_t(X)=\mathcal{E}(j_t(X)|Y(s\le t))$, where here
$\mathcal{E}(\cdot|\cdot)$ is a proper infinite-dimensional generalization
of (\ref{eq:qcondex}).  A crucial point is that $j_t(X)$ and $Y(s)$ can in
fact be shown to commute for all $s\le t$; this is called the {\it
nondemolition property} by Belavkin \cite{s:belavkin}.  Thus we see that,
even though we can evidently interpret the quantum filter in terms of the
projection postulate, we do not need to postulate anything beyond the
standard formalism of observables and expectations in quantum mechanics.  
Instead, we see that the filter follows naturally from a statistical
inference procedure wherein we find the least-squares estimate for every
system observable given the observations.  This point of view is 
very natural in a control-theoretic context.

We now give the quantum filter for our model (\ref{eq:sys}), (\ref{eq:obs});
we refer to \cite{s:VanHandel2005,s:belavkin,s:belavkz1,s:belavkz2,s:bouten}
for various approaches to deriving this equation.  The result is
\begin{eqnarray}
	d\pi_t(X)=
	\pi_t(
		(\gamma/2)\{
			\overline{\mathcal D}[\mu]+
			\overline{\mathcal D}[\mu_z]+
			\overline{\mathcal D}[\mu^\dag]
		\}X
	)\,dt
\nonumber\\
\phantom{ d\pi_t(X)=\pi}
	+\pi_t(
		2\kappa\,\overline{\mathcal D}[a]X
	)\,dt
	+\pi_t(
		i(g/2)[\mu_zx,X]
	)\,dt
\nonumber\\
\phantom{ d\pi_t(X)=\pi}
	+ \sqrt{2\kappa\eta}\,[
		i\pi_t(a^\dag X-Xa)
	-\pi_t(y)\pi_t(X)]\times
\nonumber\\
\label{eq:qfilt}
\phantom{ d\pi_t(X)=\pi + \sqrt{2\kappa\eta}\,[i\pi_t(a^\dag X-X}
	(dY(t)-\sqrt{2\kappa\eta}\,\pi_t(y)\,dt)
\end{eqnarray}
where $\overline{\mathcal D}[c]X=c^\dag Xc-(c^\dag cX+Xc^\dag c)/2$.
The process $dW(t)=dY(t)-\sqrt{2\kappa\eta}\,\pi_t(y)\,dt$ is known as the 
{\it innovations process}; it describes how ``surprised'' we are by the 
measurement, as it is the difference between the observation $dY(t)$ and 
our best estimate of what we should observe.  It can be shown that, as 
long as the observation process $Y(t)$ has the statistics determined by 
(\ref{eq:sys}) and (\ref{eq:obs}), the innovations process $dW_t$ is a 
Wiener process.  In some sense this reflects the optimality of the filter, 
as it means that the innovation is unbiased.

Usually the quantum filter (\ref{eq:qfilt}) is written in its density 
form.  To do this, we define a random density operator $\rho_t$ such 
that\footnote{
	We mean this in the sense of random variables; that is, 
	${\rm Tr}[X\rho_t]$ is a classical random variable with the same 
	statistics as the observable $\pi_t(X)$.  We have already implied
	such a correspondence by interpreting $Y(t)$ as a classical
	stochastic process.  In general, we can always express a set of
	observables as classical random variables as long as they commute 
	\cite{s:maassen}.
} $\pi_t(X)={\rm Tr}[X\rho_t]$.  We then find
\begin{eqnarray}
	d\rho_t=-i(g/2)[\mu_zx,\rho_t]\,dt
		+2\kappa\mathcal{D}[a]\rho_t\,dt
\nonumber\\
\phantom{d\rho_t=-i}
		+(\gamma/2)\{\mathcal{D}[\mu]
			+\mathcal{D}[\mu_z]+\mathcal{D}[\mu^\dag]\}\rho_t\,dt
\label{eq:dfilt}\\
\nonumber
\phantom{d\rho_t=-i}
	+\sqrt{2\kappa\eta}\,[
		i\rho_t a^\dag-ia\rho_t-{\rm Tr}[\rho_t y]\rho_t
	](dY(t)-\sqrt{2\kappa\eta}\,{\rm Tr}[\rho_t y]\,dt)
\end{eqnarray}
where $\mathcal{D}[c]\rho=c\rho c^\dag-(c^\dag c\rho+\rho c^\dag c)/2$.
This description is more economical than the raw filter (\ref{eq:qfilt}), 
and appears frequently in the physics literature.  We have to be careful, 
however, to interpret $\rho_t$ as the information state of an observer 
with access to $Y(t)$, and {\it not} as the physical state of the system.  
This point will be important for the interpretation of our results.

We conclude this section with one more filter, the so-called unnormalized 
filter, which is given by the expression
\begin{eqnarray}
	d\tilde\rho_t=-i(g/2)[\mu_zx,\tilde\rho_t]\,dt
		+2\kappa\mathcal{D}[a]\tilde\rho_t\,dt
\nonumber\\
\phantom{d\rho_t=-i}
		+(\gamma/2)\{\mathcal{D}[\mu]
			+\mathcal{D}[\mu_z]+\mathcal{D}[\mu^\dag]\}
		\tilde\rho_t\,dt
\label{eq:ufilt}\\
\nonumber
\phantom{d\rho_t=-i}
	+i\sqrt{2\kappa\eta}\,[\tilde\rho_t a^\dag-a\tilde\rho_t]\,dY(t)
\end{eqnarray}
The information state $\tilde\rho_t$ propagated by this filter is not 
normalized, ${\rm Tr}[\tilde\rho_t]\ne 1$.  However, it is simply related 
to the normalized information state by $\rho_t=\tilde\rho_t/{\rm 
Tr}[\tilde\rho_t]$.  The chief advantage of (\ref{eq:ufilt}) is that it is 
a linear equation, whereas (\ref{eq:dfilt}) is nonlinear in $\rho_t$.  
This makes (\ref{eq:ufilt}) somewhat easier to manipulate.

\subsection{The $Q$-filter}

In \cite{s:mabuchi}, it was noticed that density operators of the form
\begin{equation}
\label{eq:qansatz}
	\rho=\sum_{a=\pm}|a\rangle\langle a|\otimes
		\int dy\, P^a(y)|iy/2\rangle\langle iy/2|
\end{equation}
[$|iy/2\rangle$ are coherent states of the cavity mode] form an invariant 
set of the filtering equation (\ref{eq:dfilt}).  Thus, as long as the 
initial density is within this set, we can represent the filtering 
equations in terms of the pair of real (Glauber-Sudarshan) functions 
$P^\pm(y)$ on a line.  Substituting (\ref{eq:qansatz}) into 
(\ref{eq:ufilt}) yields the unnormalized $P$-filter
\begin{eqnarray}
	dP^\pm_t(y)=
	\frac{\partial}{\partial y}[(\pm g+\kappa y)P^\pm_t(y)]\,dt
\nonumber\\
\label{eq:pfiltt}
\phantom{dP^\pm_t(y)=(\pm g+}
	+\frac{\gamma}{2}[P^\mp_t(y)-P^\pm_t(y)]\,dt
	+\sqrt{2\kappa\eta}\,yP^\pm_t(y)\,dY(t)
\end{eqnarray}
For our purposes, it is more convenient to work with unnormalized 
$Q$-functions
\begin{equation}
\label{eq:qfunction}
	Q^\pm(y)=\langle\pm,iy/2|\rho|\pm,iy/2\rangle=
	\int dy'\, P^\pm(y')\, e^{-(y-y')^2/4}
\end{equation}
as these are always guaranteed to be well-behaved densities \cite{s:mandel}.
We obtain
\begin{eqnarray}
	dQ^\pm_t(y)=
	\frac{\gamma}{2}[Q^\mp_t(y)-Q^\pm_t(y)]\,dt
	+\frac{\partial}{\partial y}[(\pm g+\kappa y)Q^\pm_t(y)]\,dt
\nonumber\\
\label{eq:qfiltt}
\phantom{dQ^\pm_t(y)=Q}
	+2\kappa\frac{\partial^2}{\partial y^2}Q^\pm_t(y)\,dt
	+\sqrt{2\kappa\eta}\,\left[
		y+2\frac{\partial}{\partial y}
	\right]Q^\pm_t(y)\,dY(t)
\end{eqnarray}
The simplicity of this expression motivates our choice of this system for 
demonstrating the quantum projection filter.

Rather than using the $Q$-function for the projection filter, we could
work directly with the filter (\ref{eq:dfilt}) in density form and apply
methods of quantum information geometry \cite{s:amari}.  However, note
that any metric on a manifold of densities induces a metric on the
corresponding manifold of density operators (e.g.\ \cite{s:topology}).  
Thus even the $Q$-function projection filter is a true quantum projection
filter, as long as we project onto a family of $Q$-functions that
correspond to valid quantum states.

\subsection{Observing the spontaneous emission}

Until now we have only observed the forward channel; however, at least in
principle, we could also observe independently the three spontaneous
emission channels $B_{s,z}$, $B_{s,\pm}$.  We would like to identify a
spontaneous emission event with the detection of a photon in one of these
side channels.  As such, in this section we discuss the situation wherein
direct photodetection is performed in each of the spontaneous emission
channels, in addition to the homodyne detection of the forward channel.

The analysis in this case is very similar to the one performed in sections 
\ref{sec:jc1}--\ref{sec:jc2}.  The system model is still given by 
(\ref{eq:sys}).  Now, in addition to (\ref{eq:obs}), we need to introduce 
three observation processes $N_{z,+,-}$ corresponding to photodetection 
(with perfect efficiency) in the three spontaneous emission channels.
The details of this setup and the associated filtering equations are well 
known and we will not repeat them here (see e.g.\ \cite{s:bouten}).  The 
full (normalized) filtering equation is given by
\begin{eqnarray}
	d\rho_t=-i(g/2)[\mu_zx,\rho_t]\,dt
		+2\kappa\mathcal{D}[a]\rho_t\,dt
\nonumber\\
\phantom{d\rho_t=-i}
		+(\gamma/2)\{\mathcal{D}[\mu]
			+\mathcal{D}[\mu_z]+\mathcal{D}[\mu^\dag]\}\rho_t\,dt
\nonumber\\
\phantom{d\rho_t=-i}
	+\sqrt{2\kappa\eta}\,[
		i\rho_t a^\dag-ia\rho_t-{\rm Tr}[\rho_t y]\rho_t
	](dY(t)-\sqrt{2\kappa\eta}\,{\rm Tr}[\rho_t y]\,dt)
\nonumber\\
\phantom{d\rho_t=-i}
	+\mathcal{G}[\mu]\rho_t\,(dN_+(t)-(\gamma/2)\,{\rm Tr}[\mu^\dag\mu\rho_t]\,dt)
\nonumber\\
\phantom{d\rho_t=-i}
	+\mathcal{G}[\mu_z]\rho_t\,(dN_z(t)-(\gamma/2)\,dt)
\nonumber\\
\label{eq:fullfilt}
\phantom{d\rho_t=-i}
	+\mathcal{G}[\mu^\dag]\rho_t\,(dN_-(t)-(\gamma/2)\,{\rm Tr}[\mu\mu^\dag\rho_t]\,dt)
\end{eqnarray}
where $\mathcal{G}[c]\rho=c\rho c^\dag/{\rm Tr}[c\rho c^\dag]-\rho$.  It 
can be shown that the statistics of the processes $N_{+,z,-}$(t) is such 
that they are counting processes with independent jumps and rates 
$(\gamma/2)\,{\rm Tr}[\mu^\dag\mu\rho_t]$, $(\gamma/2)$ and 
$(\gamma/2)\,{\rm Tr}[\mu\mu^\dag\rho_t]$, respectively.

We now have two different filters, equations (\ref{eq:dfilt}) and
(\ref{eq:fullfilt}), for the same physical system (\ref{eq:sys}).  To see 
how they relate, recall that all the filter is propagating is an 
information state.  The information state in (\ref{eq:dfilt}) represents 
the best estimate of an observer who only has access to the homodyne 
measurement in the forward channel.  The information state in 
(\ref{eq:fullfilt}), however, represents the best estimate of a different 
observer who has access to both the homodyne observation and to direct 
photodetection of the spontaneous emission channels.  Neither information 
state represents the physical state of the system; the latter is given by 
(\ref{eq:sys}).

In practice, the frequency-resolved monitoring of spontaneously emitted 
photons is not (yet) experimentally feasible.  Hence we would never use 
the filter (\ref{eq:fullfilt}) in an actual experimental situation.  On 
the other hand, we are able to generate photocurrents $Y(t)$, 
$N_{+,z,-}(t)$ with the correct statistics in a computer simulation.  It 
is then interesting to compare the estimate of an observer who has access 
to all photocurrents to the estimate of a realistic observer who only has 
access to the forward channel.  In particular, this gives insight into the 
question asked in \cite{s:mabuchi}, `in what sense should we be able to 
associate observed phase-switching events [in the forward channel] with 
``actual'' atomic decays?'

The main reason for introducing (\ref{eq:fullfilt}) is that it gives us a
convenient way to perform computer simulations of the photocurrent $Y(t)$.  
We wish to generate sample paths of $Y(t)$, with the correct statistics,
in order to compare the performance of the optimal filter (\ref{eq:dfilt})  
with the projection filter that we will derive shortly.  Ideally we would
directly simulate the system evolution (\ref{eq:sys}); this problem is
essentially intractable, however.  Fortunately, we have already expressed
the statistics of the photocurrents $Y(t)$, $N_{+,z,-}(t)$ completely in
terms of the information state.  Hence we can equivalently simulate the
photocurrents by simulating (\ref{eq:fullfilt}) according to these rules
($dW(t)$ is a Wiener process, $N_+(t)$ has rate $(\gamma/2)\,{\rm
Tr}[\mu^\dag\mu\rho_t]$, etc.)

Of course, we can also perform such simulations with (\ref{eq:dfilt}).  
However, the advantage of (\ref{eq:fullfilt}) is that, if we choose 
$\eta=1$, the pure states are an invariant set of this filter.  We can 
thus rewrite the equation as a stochastic Schr{\"o}dinger equation, in 
which we only have to propagate a vector instead of an operator.  This is 
a much more efficient numerical procedure, and is frequently used in 
quantum optics \cite{s:gardinerparkins}.  For our system, the stochastic 
Schr{\"o}dinger equation corresponding to (\ref{eq:fullfilt}) is given by
\begin{eqnarray}
	d|\psi_t\rangle=[
		(-i(g/2)\mu_z x
		 -i\kappa\langle y\rangle_t a
		 -\kappa a^\dag a
		 -(\kappa/4)\langle y\rangle_t^2)\,dt
\nonumber\\
\phantom{d|\psi_t\rangle=[(-i}
		-\sqrt{2\kappa}\,(ia+\langle y\rangle_t/2)\,dW(t)
		+(\mu/\langle\mu^\dag\mu\rangle_t^{1/2}-1)\,dN_+(t)
\nonumber\\
\phantom{d|\psi_t\rangle=[(-i}
\label{eq:sse}
		+(\mu_z-1)\,dN_z(t)
		+(\mu^\dag/\langle\mu\mu^\dag\rangle_t^{1/2}-1)\,dN_-(t)
	]\,|\psi_t\rangle
\end{eqnarray}
where $\langle c\rangle_t=\langle\psi_t|c|\psi_t\rangle$, and 
$\rho_t=|\psi_t\rangle\langle\psi_t|$.  We numerically solve this equation 
in a truncated Fock basis for the cavity mode.  The homodyne photocurrent 
(\ref{eq:obs}) is calculated from the innovation using 
$dY(t)=\sqrt{2\kappa\eta}\,\langle y\rangle_t\,dt+\sqrt{\eta}\,dW_t+\sqrt{1-\eta}\,dV_t$.

\section{The quantum projection filter}
\label{sec:proj2}

\subsection{The finite-dimensional family}

Before we can obtain a projection filter for (\ref{eq:qfiltt}), we must 
fix the finite-dimensional family of densities to project onto.  Note that 
each density is actually the pair of $Q$-functions $Q^\pm(y)$, unlike 
in section \ref{sec:proj} where each density was a single function.  
However, we can easily put the problem into this form by making $\pm$ 
an argument of the function, i.e.\ $Q^\pm(y)=Q(y,\pm)$.  The square roots 
of $Q$-functions form a perfectly reasonable $L^2$ space 
($L^2=L^2(\mathbb{R})\oplus L^2(\mathbb{R})$) with the inner product
\begin{equation}
	\langle Q_1^{1/2},Q_2^{1/2}\rangle=
		\sum_{a=\pm}\int_{-\infty}^\infty dy\,
			Q_1^{1/2}(y,a)Q_2^{1/2}(y,a)
\end{equation}
In the following we will use the notations $Q^\pm(y)$ and $Q(y,\pm)$ 
interchangeably.

Numerical simulations of (\ref{eq:qfiltt}) show that at any time, both 
$Q^+(y)$ and $Q^-(y)$ are unimodal, roughly bell-shaped densities with an 
approximately constant width.  This suggests that we can attempt to 
approximate the information state by unnormalized density operators of the 
form
\begin{equation}
	\rho=
		\nu^+|+\rangle\langle+|\otimes
			|i\mu^+/2\rangle\langle i\mu^+/2|+
		\nu^-|-\rangle\langle-|\otimes
			|i\mu^-/2\rangle\langle i\mu^-/2|
\end{equation}
This corresponds to the bi-Gaussian family of unnormalized $Q$-functions
\begin{equation}
	q(y,\pm)=\frac{\nu^\pm}{2\sqrt{\pi}}
		\exp\left[
			-\frac{(y-\mu^\pm)^2}{4}
		\right],~\mu^\pm\in\mathbb{R},~\nu^\pm\ge 0
\end{equation}
We collect the parameters into a vector 
$\theta=(\mu^+,\nu^+,\mu^-,\nu^-)$, where 
$\theta\in\Theta=\{\mu^\pm\in\mathbb{R},~\nu^\pm\ge 0\}$.  Then the family 
of square roots of densities
\begin{equation}
	S^{1/2}=\{\sqrt{q(y,\pm;\theta)},~\theta\in\Theta\}
\end{equation}
is a finite-dimensional manifold in $L^2$ with the tangent space
\begin{equation}
	T_\theta S^{1/2}=\mbox{Span}\left\{
	\frac{\partial\sqrt{q(y,\pm;\theta)}}{\partial\theta^i}:
	i=1\ldots 4\right\}
\end{equation}
and Fisher metric
\begin{equation}
	g_{ij}(\theta)=
	4\left\langle
	\frac{\partial\sqrt{q(y,\pm;\theta)}}{\partial\theta^i},
	\frac{\partial\sqrt{q(y,\pm;\theta)}}{\partial\theta^j}
	\right\rangle
\end{equation}
Calculating the latter explicitly, we obtain the diagonal matrix
\begin{equation}
	g(\theta)=\mbox{diag}\left\{
		\frac{\nu^+}{2},~\frac{1}{\nu^+},~
		\frac{\nu^-}{2},~\frac{1}{\nu^-}
	\right\}
\end{equation}

\subsection{The projection filter}

We will perform projection of the unnormalized filtering equation 
(\ref{eq:qfiltt}), as in \cite{s:vellekoop}.  We begin by converting the 
equation into the Stratonovich form:
\begin{eqnarray}
	dQ^\pm_t(y)=
	\frac{\gamma}{2}[Q^\mp_t(y)-Q^\pm_t(y)]\,dt
	+\frac{\partial}{\partial y}[(\pm g+\kappa(1-4\eta)y)Q^\pm_t(y)]\,dt
\nonumber\\
\phantom{dQ^\pm_t(y)=Q}
	+2\kappa(1-2\eta)\frac{\partial^2}{\partial y^2}Q^\pm_t(y)\,dt
	+\kappa\eta(2-y^2)Q^\pm_t(y)\,dt
\nonumber\\
\label{eq:qfilttst}
\phantom{dQ^\pm_t(y)=Q+2\kappa(1-2\eta)\frac{\partial}{\partial y}}
	+\sqrt{2\kappa\eta}\,\left[
		y+2\frac{\partial}{\partial y}
	\right]Q^\pm_t(y)\circ dY(t)
\end{eqnarray}
We can now use (\ref{eq:projf}) and (\ref{eq:projf2}) to find dynamical 
equations for the projection filter.  After tedious but straightforward 
calculations, we obtain
\begin{eqnarray}
\label{eq:projfs1}
        d\nu^+_t = \left[\frac{\gamma}{2}(\nu^-_t - \nu^+_t)
                -\kappa\eta\,(\mu^+_t)^2\nu^+_t
        \right]dt
        +\sqrt{2\kappa\eta}\,\mu^+_t\nu^+_t\circ dY(t) \\
\label{eq:projfs2}
        d\nu^-_t = \left[\frac{\gamma}{2}(\nu^+_t - \nu^-_t)
                -\kappa\eta\,(\mu^-_t)^2\nu^-_t
        \right]dt
        +\sqrt{2\kappa\eta}\,\mu^-_t\nu^-_t\circ dY(t) \\
        \frac{d\mu^+_t}{dt} = -g-\kappa\mu^+_t+
                \frac{\gamma}{2}\frac{\nu^-_t}{\nu^+_t}
                (\mu^-_t-\mu^+_t) \\
        \frac{d\mu^-_t}{dt} = +g-\kappa\mu^-_t+
                \frac{\gamma}{2}\frac{\nu^+_t}{\nu^-_t}
                (\mu^+_t-\mu^-_t)
\end{eqnarray}
Conversion to the It\^o form changes (\ref{eq:projfs1}) and 
(\ref{eq:projfs2}) to
\begin{eqnarray}
        d\nu^+_t = \frac{\gamma}{2}(\nu^-_t - \nu^+_t)\,dt
        +\sqrt{2\kappa\eta}\,\mu^+_t\nu^+_t\, dY(t) \\
        d\nu^-_t = \frac{\gamma}{2}(\nu^+_t - \nu^-_t)\,dt
        +\sqrt{2\kappa\eta}\,\mu^-_t\nu^-_t\, dY(t)
\end{eqnarray}
Finally, we rewrite the equations in terms of the {\it normalized} 
parameters $\mu^\pm$ and $\tilde\nu_t^+=\nu_t^+/(\nu_t^++\nu_t^-)$.  This 
gives
\begin{eqnarray}
        d\tilde\nu^+_t = -\gamma(\tilde\nu^+_t-1/2)\,dt+
        \sqrt{2\kappa\eta}\,\tilde\nu^+_t(1-\tilde\nu^+_t)
                (\mu^+_t-\mu^-_t)\times
\nonumber\\
\label{eq:pf1}
\phantom{d\tilde\nu^+_t = -\gamma(\tilde\nu^+_t-1/2)}
        \{dY(t)-\sqrt{2\kappa\eta}\,[\mu^+_t\tilde\nu^+_t
                +\mu^-_t(1-\tilde\nu^+_t)]\,dt\} \\
\label{eq:pf2}
        \frac{d\mu^+_t}{dt} = -g-\kappa\mu^+_t+
                \frac{\gamma}{2}\frac{1-\tilde\nu^+_t}{\tilde\nu^+_t}
                (\mu^-_t-\mu^+_t) \\
\label{eq:pf3}
        \frac{d\mu^-_t}{dt} = +g-\kappa\mu^-_t+
                \frac{\gamma}{2}\frac{\tilde\nu^+_t}{1-\tilde\nu^+_t}
                (\mu^+_t-\mu^-_t)
\end{eqnarray}
Equations (\ref{eq:pf1})--(\ref{eq:pf3}) form the projection filter 
for our model on the family $S^{1/2}$.

Note that equations (\ref{eq:pf2}) and (\ref{eq:pf3}) are singular at 
$\tilde\nu^+_t=0$ or $1$.  We can trace this back to the fact that we have 
cheated a little in the definition of our family of densities. 
When $\nu^+=0$ (or $\nu^-=0$), the map $\theta\mapsto q(y,\pm;\theta)$ is 
not invertible, as in this case any choice of $\mu^+$ (or $\mu^-$) leads 
to the same density.  As we have essentially inverted this map to obtain 
the equations (\ref{eq:pf1})--(\ref{eq:pf3}) for the parameters, we can 
hardly expect these to be well-defined when this map is not invertible.  

Fortunately the points $\tilde\nu^+_t=0$ and $1$ are never reached if 
we start the filter with $0<\tilde\nu^+<1$.  Hence we can make the filter 
well-defined everywhere simply by removing the offending points 
$\nu^+_t=0$ and $\nu^-=0$ from $S^{1/2}$.  The map $\theta\mapsto 
q(y,\pm;\theta)$ is then invertible everywhere (in other words, then the 
manifold is covered by a single chart.)  Even if we want to consider 
starting the filter on $\tilde\nu^+_t=0$ or $1$ at $t=0$ this is not a 
problem; the filter dynamics will cause $\tilde\nu^+$ to evolve off the 
singular point, so that the filter is well defined after an arbitrarily 
small time step \cite{s:vellekoop}.

\subsection{Connection with the Wonham filter}

There is a remarkable connection between the projection filter obtained in 
the previous section and the theory of jump process filtering.  This 
theory goes back to the beautiful classic paper by Wonham \cite{s:wonham}, 
in which the following problem is solved.

Denote by $x(t)$ a stationary Markovian jump process which switches 
between two states $a_-$ and $a_+$ with a rate $\gamma/2$; i.e., $x(t)$ is 
a random telegraph signal \cite{s:gardinerhb}.  Now suppose we do not have 
access to a complete obeservation of $x(t)$, but only to the corrupted 
observation $y(t)$ defined by
\begin{equation}
\label{eq:wonhamobs}
	dy(t)=\sqrt{2\kappa\eta}\,x(t)\,dt+dw(t)
\end{equation}
where $dw(t)$ is a Wiener process.  We can now ask, what is our best guess 
of the probability $p_+(t)$ that $x(t)=a_+$, given the observations 
$y(s\le t)$?  The answer is given in closed form by (a special case of) 
the Wonham filter:
\begin{eqnarray}
	dp_+(t)=-\gamma[p_+(t)-1/2]\,dt+
		\sqrt{2\kappa\eta}\,p_+(t)[1-p_+(t)](a_+-a_-)\times
\nonumber\\
\phantom{dp_+(t)=-\gamma[p_+(t}
	\{dy(t)-\sqrt{2\kappa\eta}\,[a_+p_+(t)+a_-(1-p_+(t))]\,dt\}
\end{eqnarray}
But this is exactly (\ref{eq:pf1}) with $\mu^+_t$ and $\mu^-_t$ replaced 
by the constants $a_+$ and $a_-$.

Though intuitively appealing, this is in many ways a remarkable result.
There appear to be no inherent jumps in either the optimal filter
(\ref{eq:dfilt}), or the system-observation pair (\ref{eq:sys}),
(\ref{eq:obs}) from which it was obtained.  It is true that we can choose
to observe a jump process in the spontaneous emission channels, as in
(\ref{eq:fullfilt}), but we could have equally chosen to perform homodyne
or heterodyne detection which do not lead to jump process observations.
Nonetheless (\ref{eq:pf1}) emerges naturally from our model, and an
expression of the same form can even be obtained directly from
(\ref{eq:dfilt}) \cite{s:mabuchi}.  Evidently there is a deep connection
between our system and the theory of jump processes.

As a classical filter, we can interpret the projection filter
(\ref{eq:pf1})--(\ref{eq:pf3}) as an {\it adaptive} Wonham filter, where
the equations (\ref{eq:pf2})--(\ref{eq:pf3}) continually adapt the
parameters $a_+$ and $a_-$ in the Wonham filter (\ref{eq:pf1}).  A similar
structure was observed in \cite{s:vellekoop}, where the classical problem
of changepoint detection (the detection of a single random jump in white
noise) was treated using the projection filtering approach.

\section{Numerical results}
\label{sec:num}

\begin{figure} 
\includegraphics[width=\textwidth]{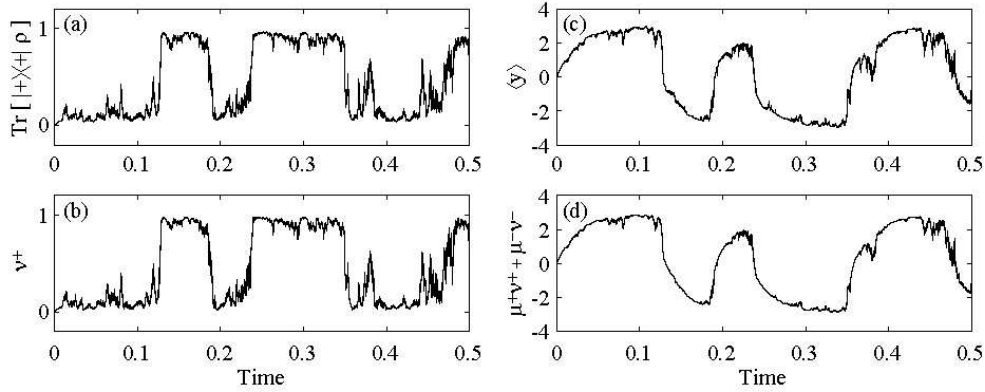} 
\caption{
\label{f:compare} Comparison between the optimal and projection filters.  
A typical observation is shown with $\eta=1$, $g=120$, $\kappa=40$,
$\gamma=20$, and the integration was performed over $25\,000$ time steps.
The top row, figures (a) and (c), were calculated using the optimal filter 
(\ref{eq:qfiltt}); the bottom row, (b) and (d), were calculated for the 
same observation process using the projection filter 
(\ref{eq:pf1})--(\ref{eq:pf3}).}
\end{figure}

In this section we present the results of numerical simulations of the
various filters.  Sample paths of the observation process were generated
by numerically solving (\ref{eq:sse}) with a truncated cavity basis of
$25$ Fock states and the (appropriately truncated) initial state
$|\psi_0\rangle=|-\rangle\otimes|0\rangle$. The thus generated
observations were then filtered using the optimal filter in $Q$-function
form (\ref{eq:qfiltt}), and using the projection filter
(\ref{eq:pf1})--(\ref{eq:pf3}).

The optimal filter was implemented using a simple finite-difference scheme
\cite{s:nr} on a grid of $128$ equidistant points in the interval
$y\in[-18,18]$, with the appropriately truncated initial condition
corresponding to $|\psi_0\rangle$.  Finally, the projection filter was
started with the initial condition $\tilde\nu^+=\mu^+=\mu^-=0$, where care
was taken not to propagate $\mu^+$ until after the first time step.  In 
all simulations, stochastic integration was performed using the stochastic 
Euler method \cite{s:kloeden}.

In Figure \ref{f:compare} a typical filtered sample path is shown.  The 
top row was obtained from the optimal filter, while the bottom row was 
obtained using the projection filter.  The value inferred for both the 
conditional probability of finding the atom in the $|+\rangle$ state (left 
column) and the conditional expectation of the $y$-quadrature (right 
column) are nearly identical for the two filters.  Evidently the 
projection filter is an extremely good approximation to the optimal 
filtering equation.

\begin{figure} 
\includegraphics[width=\textwidth]{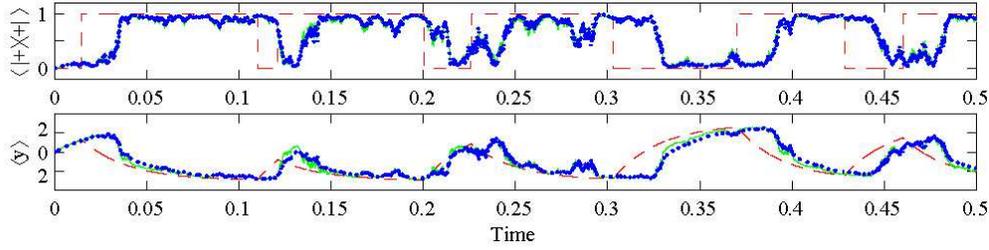} 
\caption{
\label{f:typical}  A single run of an experiment is simulated.  The dashed 
line (red) corresponds to the optimal estimate of an observer who has 
access to direct photodetection of the spontaneously emitted photons, as 
well as homodyne detection of the forward channel.  The solid line (green) 
is the optimal estimate of a different observer, who only has access to 
the homodyne photocurrent, for the same run of the experiment.  The dotted 
line (blue) is the projection filter estimate based only on the homodyne 
photocurrent.  All parameters are the same as in Figure \ref{f:compare}.}
\end{figure}

\begin{figure} 
\includegraphics[width=\textwidth]{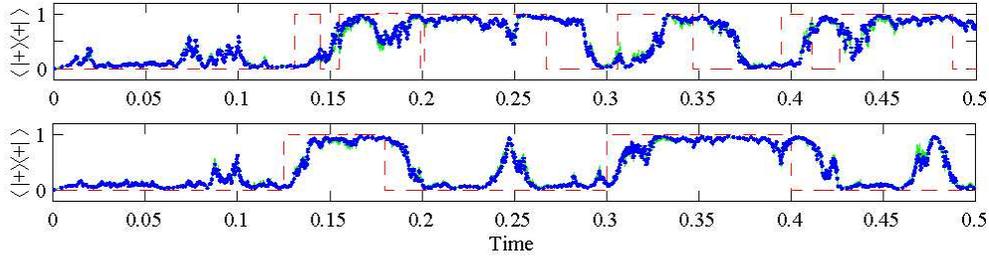} 
\caption{
\label{f:atypical}  Two more sample paths of the filtered estimate 
of atomic state, demonstrating missed jumps (top) and false jumps 
(bottom).  All parameters are the same as in Figure \ref{f:compare}.}
\end{figure}

Next, in Figure \ref{f:typical}, we compare the information state of the 
optimal and projection filters to the information state that is based on 
the additional observation of spontaneously emitted photons.  The latter 
filter demonstrates the behavior reported in \cite{s:alsing}.  Whenever a 
photon is observed in one of the side peaks of the Mollow triplet, the 
observer infers that the atom has made a jump.  The estimated phase of the 
cavity field then exponentially decays to a steady-state level of $\langle 
y\rangle=\pm g/\kappa$.

If we do not measure the spontaneous emission, our best guess of the
atomic state still behaves in a jump-like way.  However, we see that there
is a little delay between the time that the observer of spontaneous
emission thinks the atom has jumped, and the time that the homodyne
observer comes to the same conclusion.  If we identify atomic decay with
the spontaneous emission of a photon, we can now give a fairly
satisfactory answer to the question posed in \cite{s:mabuchi}: `In what
sense should we be able to associate observed phase switching events with
``actual'' atomic decays'?  It appears that an observed phase switch
signals that, had we been making such an observation, we would likely have
seen a spontaneously emitted photon a little while earlier.

The detection delay is a rather generic property of the type of filtering 
problems we are considering \cite{s:vellekoop,s:shiryaev}.  Any time we 
see a large fluctuation in the observed process, the filter has to decide 
whether this is a large fluctuation of the noise, or a large fluctuation 
of the observed system.  As is pointed out by Shiryaev \cite{s:shiryaev} 
in the context of changepoint detection, the filter rides a delicate 
balance between minimizing the delay time and minimizing the probability 
of ``false alarms''.  Decreasing the number of false alarms (by 
choosing a different filtering cost) would unavoidably increase the delay 
time, and vice versa.  In our system, false alarms are missed jumps 
and false jumps; these do occur, as can be seen in Figure 
\ref{f:atypical}.  

\begin{figure} 
\includegraphics[width=\textwidth]{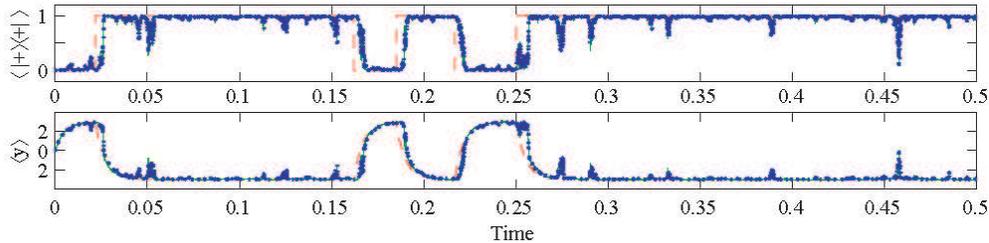} 
\caption{
A typical filtered sample path with $\eta=1$, $g=600$, $\kappa=200$,
$\gamma=20$. The integration was performed over $100\,000$ time steps.
The various line types are as in the previous figures.
\label{f:better}
}
\end{figure}

If we wish to generally improve the quality of detection we have no 
alternative than to increase the signal-to-noise level of the observation.  
In the case of our system, we can do this if we increase $g$ and $\kappa$
while keeping their ratio fixed (the analogy can be justified from 
(\ref{eq:wonhamobs}); the signal $x$ has fixed magnitude $x=\pm g/\kappa$, 
while the signal-to-noise ratio $\sim g/\sqrt{\kappa}$.)  A simulation 
with greatly increased signal-to-noise is shown in Figure \ref{f:better}.
In this very strong coupling and damping regime, it appears that not much 
more information can be extracted from observation of the spontaneous 
emission than we could have already inferred from the homodyne 
photocurrent.

\section{Conclusion}

In this paper, we have suggested that the method of projection filtering
can be very fruitful when applied to quantum filtering theory.  Using a
simple model of a strongly coupled two-level atom in a cavity we
numerically demonstrated near-optimal performance of the projection
filter, as is evident from Figures \ref{f:compare}--\ref{f:better}.  We
have also shown a connection between this model from cavity QED and the
classical Wonham filter; the projection filter can be interpreted as an
adaptive Wonham filter, applied to a quantum model.  In future work we 
will develop a ``true'' quantum formalism for projection filtering, 
using methods from quantum information geometry.

The reduction of infinite or high-dimensional filters to a tractable set 
of equations is essential if we wish to perform estimation in real time, 
for example in a feedback control loop.  In a control-theoretic context, 
converting a large, complex system into a set of simple equations is known 
as model reduction.  Ideally, however, such a procedure should yield some 
bounds on the error of approximation; in our case we have observed 
numerically that the approximation error is very small, but we have no 
rigorous bounds to back up this statement.  In classical control theory of 
linear systems, the method of balanced truncation \cite{s:dullerud} gives 
a very general method for model reduction with guaranteed error bounds.
How to do this effectively for nonlinear systems is still an open problem, 
however, both in classical and in quantum theory.

\section*{Acknowledgment}

The authors thank Mike Armen for useful discussions.  
This work was supported by the ARO (DAAD19-03-1-0073).

\section*{References}

\bibliographystyle{unsrt}
\bibliography{QPF}

\end{document}